
\def\sect{\vskip 2mm \centerline}

\def\r{\hangindent=1pc  \noindent}
\def\ref{\hangindent=1pc  \noindent}

\def\v{\vskip 1mm}

\def\endpage{\vfil\break}

\def\noi{\noindent}
\def\kms{km s$^{-1}$}

\def\deg{$^\circ$}

\def\Msun{M_{\odot \hskip-5.2pt \bullet}}

\def\Deg{^\circ}
\def\deg{$^\circ$}

\def\microg{$\mu$G}

\def\kluwer{Kluwer Academic Publishers, Dordrecht}

\def\so{Sofue, Y.}

\hsize=128truemm
\vsize=196truemm
\def\sect{\v\noi}
\noi Presented at a NATO adv. School ``The Nuclei of Normal Galaxies: Lessons
from the Galactic Center", Ringberg Schloss, Tegernsee, 1993 July 25 - 30.
\topskip 10truemm
\noindent {\bf LARGE-SCALE RADIO STRUCTUERS IN THE GALACTIC CENTER}

\vskip 15truemm

\leftskip 20mm

\noi Yoshiaki SOFUE

\noi {\it Institute of Astronomy, University of Tokyo}

\noi {\it Mitaka, Tokyo 181, Japan}

\noi sofue@sof.ioa.s.u-tokyo.ac.jp

\vskip 10truemm

\leftskip 0mm

\noindent {\bf Abstract.}  Radio continuum observations of the galactic center
region have revealed a number of vertical structures running across the
galactic plane. Most of the vertical structures are reasonably attributed
either to poloidal magnetic field or to energy release toward the halo.
The relation of the continuum structures to the molecular gas rings and their
vertical extension are also discussed.
Some large-scale ejection features appear to have similarity to bubbles found
in external galaxies.

\vskip 5truemm
\noindent {\bf Keywords}: Galactic center;  Radio emission; Jets; Magnetic
field

\topskip 0pt

\vskip 10truemm

\noi{\bf 1. Introduction}
\vskip 2mm

Radio continuum features in the galactic center region are a superposition of
star-forming regions, which comprises a thin thermal disk, and vertical
structures, which are mostly nonthermal closely related poloidal magnetic
fields.
The radio emission is therefore a mixture of thermal and nonthermal emissions.
In this paper we review radio continuum observations of the galactic center
with a particular regard to large-scale  extended features (larger than a few
tens of pc).
We first discuss the properties of the radio emission, reviewing the methods to
distinguish thermal and nonthermal emissions.
We then summarize various exotic structures, which are mostly perpendicular to
the galactic plane, and discuss them in relation to vertical magnetic fields
and to manifestation of energy release  from the nuclear disk.
We comment on the similarity of large-scale ejection features to some radio
bubble features in external galaxies.
Vertical cylinder structures of expanding rings of molecular gas are also
discussed.

The galactic center region has been mapped in the radio continuum at various
frequencies:
some of typical surveys have been made at 80 MHz (LaRosa and Kassim 1985); 160
MHz (Dulk and Slee 1974; Yusef-Zadeh et al); 327 MHz (Yusef-Zadeh et al); 408
MHz (Little 1974; Haslam et al 1982); 610 MHz (Downes et al 1978); 843 MHz
(Mills and Drinkwater 1984); 1.4 GHz (Yusef-Zadeh et al 1984; Liszt 1986; Reich
et al 1990); 2.7 GHz (Reich et al 1984); 5 GHz (Altenhoff et al 1979); 10 GHz
(Pauls et al 1976; Handa et al 1987; Seiradakis et al 1985, 1989; Haynes et al
1993); 15.5 GHz (Kapitzky and Dent 1974); 32 GHz (Reich 1993); 43 GHz (Sofue et
al 1986); 91 GHz (Tsuboi et al 1988).
There have been various reviews, some which are by Lo (1986); Brown and Listzt
(1984);  Genzel and Townes (1987); Sofue (1989).

\v\v
\noi{\bf 2. Radio Continuum Emission}
\vskip 2mm

The radio continuum emission from the galactic center region of $ \sim 3\Deg
\times 3\Deg$ is a mixture of nonthermal (synchrotron) and thermal (free-free)
emissions.
In order to investigate the various observed structures it is essential to
clarify their emission mechanisms.
 We describe the methods to separate these two emission components.

\sect{\it 2.1 Flat Radio Spectra}

The conventional method to investigate the emission mechanism is to study the
spectral index.
Thermal emission shows a flat spectrum of $\alpha= -0.1$ with $S \propto \nu
^{\alpha}$, while nonthermal emission usually shows steeper spectrum.
Mezger and Pauls (1979) studied radio emission from the galactic center, and
concluded that the extended component  comprises an ellipsoidal disk of diffuse
thermal gas of a size of $\sim 300 \times 100$ pc.
However, a spectral index study of radio data in large frequency separation,
from 845 MHz to 43 GHz, has shown that the spectral index in the radio Arc and
bridge region near Sgr A is flat or even inverted (positive) (Reich et al
1988).
It was shown that the spectral index in the central $3\Deg$ region is almost
everywhere flat (Sofue 1985).
Flat spectra are obtained even in regions where strong linear polarization was
detected (see Reich 1993 in this issue).

Hence, the flat spectrum observed near the galactic center must no longer be
taken as an indicator of thermal emission.
We note that many active galactic nuclei in extragalactic systems often show
flat radio spectra for their nonthermal characteristics.
However, the question why the spectra are so flat, or how high-energy electrons
are supplied so efficiently in such a wide area of the galactic center, is not
answered as yet.

\sect{\it 2.2. Linear Polarization}

A direct and more convincing way to distinguish synchrotron radiation is to
measure the linear polarization.
However, an extremely high Faraday rotation caused by the dense interstellar
matter and long path length through the galactic plane depolarizes the emission
by the finite-beam and finite-bandwidth effects, which makes the measurement
difficult.
This  difficulty has been resolved by the development of a multi-frequency,
narrow-band Faraday polarimeter (Inoue et al 1984) as well as by
high-resolution and high-frequency observations using the VLA (Yusef-Zadeh et
al 1986; Inoue et al 1988).
Very large rotation measure ($RM > \sim 10^3 $  rad m$^{-2}$)
and high degree (10 - 50\%) polarization have been observed along the radio Arc
and on the eastern ridge of the Galactic Center Lobe (Inoue et al 1984; Tsuboi
et al 1986; Seiradakis et al 1985; Sofue et al 1986; Reich 1988; Haynes et al
1992; Reich 1993).

At high frequencies the Faraday depolarization becomes less effective. Reich
(1988; 1993 in this issue) has reported the detection of polarization as high
as  $p\sim50$ \% along the Arc at mm wavelengths.
This is nearly equal to the theretical maximum, $p_{\rm max}=(\alpha
+1)/(\alpha+7/3)\simeq 47$ \%, for the Arc region where the spectral index has
the value of $\alpha\simeq+0.2$.
This fact implies that the magnetic field is almost perfectly ordered. This is
also consistent with the VLA observations showing straight filaments suggestive
of highly ordered magnetic field (Yusef-Zadeh et al 1984; Morris 1993 in this
issue) (Fig. 1).
{}From these observations of linear polarization it is clear that the radio
emission near the galactic center, in particular around the radio Arc, is
nonthermal despite their flat or inverted spectra.

\sect{\it 2.3 Infrared-to-Radio Ratio}

Separation of thermal and nonthermal radio emission can be done in a more
efficient way even for the regions where no polarization data are available.
The method uses comparison of far-IR (e.g. 60 $\mu$m  data from IRAS survey)
and radio intensities (both in Jy/str): thermal emission regions, mostly
composed of HII gas, have high IR-to-radio ratio, $R=I_{\rm FIR}/I_{\rm
R}\simeq 10^3$.
On the other hand, nonthermal emission regions, such as supernova remnants,
have small IR-to-radio ratio, $R = 0 \sim 300$.
Using this characteristics we are able to distinguish thermal and nonthermal
emission regions in a wide area near the galactic center (Reich et al 1987).
The region near the galactic plane is dominated by thermal emission and many
strong radio sources like Sgr B2 appear thermal (HII) (see, e.g., Mezger and
Pauls 1979).
These regions are closely associated with dense molecular clouds (Bally et al
1987) and therefore related to star formation from the clouds.
On the other hand we find many of the prominent features like the Radio Arc,
Sgr A and regions high above the galactic plane including the GCL (galactic
center lobe) are nonthermal.

\sect{\bf 3. Three-Dimensional Morphology}
\vskip 2mm

The radio continuum mappings have revealed various exotic features of peculiar
morphology (Fig. 1).
Many of them appear to lie not parallel to the galactic plane, but   run
perpendicular to the disk, which  are most likely related to poloidal magnetic
fields.

\sect{\it 3.1. The Thermal Disk and Filaments}

The nuclear disk about 50 pc thick and 200 pc in radius comprises numerous
clumps of thermal emission regions, most of which are active star-forming
regions and HII regions, are detected in the hydrogen recombination lines
(Mezger and Pauls 1979; Pauls et al 1979).
Typical HII regions are named Sgr B, C, D and E.
The total HII mass of $2\times10^6\Msun$ has been estimated, and the production
rate of Ly continuum photons of $3 \times 10^{52}$ s$^{-1}$ is required to
maintain this amount of HII gas (Mezger and Pauls 1979).
The star formaing rate, which is assumed to be proportional to the Ly continuum
photon flux, of the central few hundred pc region reaches almost 10\% of the
total star forming rate of the Galaxy.
The HII regions are surrounded by expanding shells and cylinders of molecular
gas (see 4.3).

Besides star formation regions, various exotic thermal structures have been
observed:
Complex filaments are extending from Sgr A toward the north, composing a bridge
between the radio Arc and Sgr A (Yusef-Zadeh et al 1984; Yusef-Zadeh and Morris
1988).
Detection  of hydrogen recombination lines (Pauls et al 1976; Yusef-Zadeh et al
1986) and the association of molecular gas at negative velocities (Serabyn and
G{\"u}sten 1986; 1987; G{\"u}sten 1989) indicate its thermal characteristics,
and the structures are often called the thermal arched filaments.

Despite of thermal radio emission, a large Faraday rotation has been detected
toward the bridge, which indicates  the existence of a magnetic field along the
thermal filaments (Sofue et al 1987).
A magneto-ionic jet model in which the gas is flowing out of Sgr A and collides
with the ambient poloidal magnetic field at the radio Arc has been proposed
(Sofue and Fujimoto 1987).
Serabyn and G{\"u}sten (1987) argue that the gas is flowing toward the center
and an accretion model has been proposed by analysing the velocity field.
The complex structure in the bridge indicates high turbulent motion inside the
bridge.
In fact, velocity dispersion as high as 30-50 \kms has been observed in the
bridge from recombination-line observations (Pauls et al 1976), and it
increases drastically near the Arc (60-70 \kms), which suggests a dynamical
interaction of the bridge with the Arc.
Yusef-Zadeh and Morris (1988), based on their high-resolution maps, argue that
the Arc (straight filaments) and the arched filaments are interacting with each
other.

\sect{\it 3.2. Vertical Magnetic Tubes: Radio Arc and Threads}

The radio Arc was originally found in the radio continuum survey maps of the
galactic plane (Downes et al 1978), and has been resolved into many straight
filaments with the use of VLA (Morris and Yusef-Zadeh 1987, 1988; Yusef-Zadeh
et al 1984; Yusef-Zadeh 1986; 1988; 1989; Morris 1993 in this issue) (Fig. 1).
The straight filaments run perpendicular to the galactic plane, and extend more
than $\sim 100$ pc toward positive latitudes.
The mid point of the arc is strongly polarized ($\sim20 - 50$ \%), and shows a
high Faraday rotation ($RM \sim $ a few $10^3 - 10^4$ rad m$^{-2}$) (Inoue et
al 1984; Tsuboi et al 1986; Sofue et al 1987; Yusef-Zadeh and Morris 1988;
Reich 1988; Reich 1993).
The theoretically maximum polarization of 50\% along the Arc (Reich  1993)
indicates a highly ordered magnetic field and is consistent with the VLA
straight filaments.
The magnetic field direction, as detemined from the intrinsic polarization
angles, is parallel to the filaments and vertical to the galactic plane.
Field strength as high as $\sim 1$ mG has been estimated in the Arc and in some
radio filaments (e.g., Morris 1993).

Interferometric observations of the filaments in the Arc at 43 GHz revealed
that some filaments are not visible at this high frequency.
Assumeing the field strength of the order of 1 mG, we could estimated the life
time of cosmic-ray electrons emitting at 43 GHz to be about 4000 years (Sofue
et al. 1992).
This implies that the filaments of strong magnetic field may be a time
variable, or trasient feature, being temporary illuminated by recently
accelerated high-energy electrons.
It would be, therefore, interesting to perform a time-variation watch of the
thin nonthermal filaments in the Arc by high-resolution VLA observations.

The higher latitude extensions of the Arc, both toward positive and negative
latitudes for more than 100 pc, are also  polarized and are called polarized
plumes (Tsuboi et al 1986; Yusef-Zadeh and Morris 1988). The  degree of
polaraization reaches as high as 20\% at 10 GHz. The magnetic field directions
are parallel to the radio arc, or vertical to the galactic plane.
High Faraday rotation ($\sim \pm 10^3$ rad m$^{-2}$) has been detected (Tsuboi
et al 1986; Sofue et al 1987).
The sense of rotation measure reverses from positive to negative latitude
sides, indicating a reversal of the line sight component of the magnetic field
above and below the galactic plane. The field strength is estimated to be
10-100 \microg. Sofue et al (1986) suggest that the polarized spot and the
plumes are parts of a large-scale poloidal magnetic field twisted by the disk
rotation (see next section).

Besides the filaments in Arcs, numerous straight filaments, called ``threads'',
are observed.
They appear to be distributed rather independently of the other major radio
sources  (Morris and Yusef-Zadeh 1985; Yusef-Zadeh 1988, 1989; Morris 1993).
They are roughly perpendicular to the galactic plane.
The threads intersect the other features without  interaction.
{}From the very thin and straight appearence the threads are likely magnetic
structures, and the vertical nature is consitent with a large-scale poloidal
field in the central region (Morris 1993).

\sect{\it 3.3. Large-Scale Ejection: Lobes and Jets}

The galactic center lobe (GCL) is a loop structure of about 200 pc diameter
extending vertically over the galactic plane (Sofue and Handa 1984; Sofue 1985;
Fig. 1).
The cross section of the lobe parallel to the glactic plane indicates its
cylindrical structure.
The eastern ridge of the lobe is an extension from the radio Arc and is
strongly polarized.
The magnetic field is shown to run parallel to the ridge.
The ridge also extends toward negative latitude, where a symmetric polarized
plume is found.
The western ridge emerges from Sgr C and is clearly the extension of the VLA
filament.
Yusef-Zadeh (1988) reports the detection of filament and polarization in the
western ridge, indicating the existence of magnetic field.
Recently, Uchida et al (1993) found a complex of molecular gas and warm dust
associated with the western ridge of the lobe.
They identified a velocity jump in the CO-line spectra with a shock front
coinciding with the western ridge.
By pointing out that the western ridge has a different characteristics from
that of the eastern ridge, Uchida et al (1993) suggest that both ridges might
be independent objects.

Formation of the lobe structure has been moldeled in various ways:
An explosion hypothesis requires an explosive energy injection near the
nucleus, which produces an expanding propagating the disk and halo (Sofue
1984).
An MHD acceleration model in which the gas is accelerated via a twist of
poloidal magnetic field by the accreting gas disk has been proposed (Uchida et
al 1985; Uchida and Shibata 1986).

On a much larger scale of a few kpc above the galactic plane, a giant spur has
been found, which emanates from the galactic center toward positive latitude,
$b\sim 25\Deg$ (Sofue et al 1988) (Fig. 2).
A ridge connecting the spur to the galactic center has been detected by 1408
MHz observations (Sofue et al 1988).
This feature, which is 4-kpc long and some 200-pc in diameter, may be
cylindrical in shape and extends roughly perpendicular to the galactic plane.
This structure might be a jet (or the remnant of a relativistic beam from the
nucleus), or it might be magnetic tornado produced by the differential rotation
between the halo and the nuclear disk.
Recently, Krichbaum et al (1993) found a VLBI mini-jet in the central few mas
region of Sgr A$^*$ at 43 GHz.
The jet direction appears to point the largest scale jet at high latitude, and
they have suggested a possible connection of both objects.
Falcke et al (1993) modeled the mini jet and suggested that the 4 kpc-scale jet
might be a smoke of a past similar jet phenomenon at the nucleus.

\sect{\it 3.4. The North Polar Spur and its similarity to Extragalactic
Bubbles}

The whole sky radio map at 408 MHz (Haslam et al. 1982) shows numerous radio
spurs.
The most prominent spur is called the North Polar Spur (NPS), which traces a
giant loop on the sky of diamter about 120\deg, drawing a huge $\Omega$ over
the galactic center (Fig. 2).
There have been various interpretations of this prominent  feature,
particularly related to a nearby supernova remnant.
Here, we comment on a rather exotic idea, which presumes an explosion at the
galactic center:
The giant  $\Omega$-shape of the NPS could be a shock front due to a gigantic
explosion or a sudden energy input at the galactic center of the order of
$10^{55} \sim 10^{56}$ erg (Sofue 1984). In this model, the NPS must lie at a
distance of a few kpc away.
On the other hand, the current supernova remnant hypoethesis predicts a
distance of a few tens of pc.
A key to explore the distance is the X-ray absorption by the galactic disk.
If it is a galactic-scale bubble beyond the HI disk, the X rays must be
absorbed near the galactic plane: the optical depth (H column density) changes
as a function of cosec $b$.
On the other hand, if the NPS is a local object embedded in the galactic HI
disk, the optical depth does not change with $b$, and hence, the NPS ridge
should be visible even near the galactic plane at $b<10\Deg$ where the radio
brightness is the highest.
In this context, the ROSAT data are most important (Predehl 1993).

Similar huge bubbles in radio cotninuum have been found in many spiral galaxies
(Fig. 2):
Edge-on galaxy NGC 3079 has  S-shaped double lobes which extend for 3 kpc size
in both directions of the galactic plane (Duric et al 1983).
NGC 4258 is known for its symmetrical S-shaped radio features perpendicular to
the major axis (van Albada 1980).
These bubbles would look like similar to the NPS if the observer is sitting
inside the galaxies.

\sect{\bf 4. Origin of Vertical Structures}
\vskip 2mm

\sect{\it 4.1 Poloidal fields}

Polarization observations of the galactic center region show that the field
direction projected on the sky along the radio Arc and the eastern ridge of the
galactic center lobe is perpendicular to the galactic plane (Yusef-Zadeh et al
1984, 1989; Inoue et al 1984; Tsuboi et al 1986; Sofue et al 1987; Reich 1993).
Haynes et al (1992) stressed that the polarization is visible in a wider area
of $1.5\Deg \times 1.5\Deg$ area around Sgr A, and showed that magnetic field
is widely distributed in the galactic center region.
Measurements of the  Faraday rotation  show that the rotation measure (RM)
reverses from the lower side of the galactic plane to the upper side,
indicating  reversal of the line-of-sight component of the field.
The reversal of the RM is also observed across the rotation axis of the Galaxy.
{}From these, a poloidal field model has been proposed in which the field lines
are twisted by the disk rotation (Sofue and Fujimoto 1987).
However, it is often claimed that the very straight nature of the filaments in
the radio Arc is suggestive of a field lines not bending.
If this is the case, the Faraday reversal should be attributed to intervening
fields between the Arc and the Sun.
Since the  RM reversal happens in $\sim10'$-scale and the RM amplitude is as
large as $RM\sim 10^4$ rad m$^{-2}$, it is reasonable to suppose that the
Faraday effect occurs in the nuclear disk.

If we stand on the primordial origin hypothesis of galactic magnetic fields, we
should inevitably consider the evolution of the vertical component of the
primordial field. The time scale with which a magnetic field component
perpendicular to the disk plane diffuses away from the galaxy may be estimated
as  $\tau_{\rm vert}\sim r^2/lv \sim 10^{11}$ y.
Therefore, the vertical component trapped from the intergalactic space can
never escape from the galaxy, but is transferred to the central  region when
the galaxy contraction proceeds.
A rough estimation shows that the field strength of the vertical component in
the galaxy is proportional to the surface mass density (Sofue and Fujimoto
1987), and a strong vertical field, some 10 \microg, is expected to exist in
the central 1-kpc region. On the other hand, the disk component parallel to the
galactic plane annihilates in the central region more efficiently than in the
outer region and a rather weaker disk field is expected there.

This scenario of evolution of the galactic field thus predicts that a vertical
field dominates in the central region, while the disk spiral field has the
maximum in the outer disk several kpc away from the central region.
This is consistent with the fact that our Galaxy possesses a twisted poloidal
field in the center, while the disk field (parallel to the galactic plane and
spiral arms) is strongest at $r \sim 5$ kpc.
Similar characteristics have been found in spiral galaxy M31 (Berkhuijsen et al
1987): the central radio source shows  polarization consistent with the
vertical field.
The projected field direction is perpendicular to the major axis, while the
disk field, which is a superposition of ring and spiral, has the maximum at $r
\sim 10$ kpc.

Given a large-scale poloidal field, which penetrates the rotating dense nuclear
disk, twist of the field is inevitable.
The twist will then propagate toward the halo, accelerating the gas
perpendicular to the disk plane. This will result in a cylindrical outflow of
gas and cosmic rays, which will account for the formation of the galactic
center lobe (Uchida et al 1985; Uchida and Shibata 1986; Shibata and Uchida
1987).
The large-scale poloidal field in the halo will also be twisted by the
differential rotation beween the disk and halo gases.
The twist will amplify the field strength near the rotation axis, and will be
observed as the galactic center jet (or tornado ) of 4-kpc length (Sofue et al
1988).

\sect{\it 4.2. Explosion}

A sudden energy release at the galactic center such as an explosion or a star
burst also results in vertical gaseous structurs.
In fact molecular  and hydrogen line observations of the disk gas exhibit
various expanding ring features.  We here discuss the possibility that an
explosion at the center will produce the vertical structures and examine if the
expanding rings in the disk plane are associated with vertical strucures.

An explosion and an energy release in a short time like a star burst at the
center of a rotating disk can be theoretically traced by hydrodynamic and MHD
codes. A shock wave which is initially spherical expands more rapidly in the
direction perpendicular to the disk plane because of the smaller  gas pressure
and mass in this direction compared to those in the disk plane. Then the shock
front attains an $\Omega$ shape elongated perpendicular to the disk, mimicking
the GCL at some stage (Sofue 1984).
Further propagation of the shock front through the nuclear disk has been
numerically simulated:
Refraction of the shock waves due to the density variation in the disk results
in focusing onto a ring of radius about twice the scale height of the disk gas
density ($\sim 200$ pc) in the nuclear disk.
Such a ring has been indeed observed as the 200-pc expanding ring as follows.

\sect{\it 4.3. Expanding Molecular Cylinders}

Expanding ring features of various scales are found in the galactic plane,
which are mainly observed in the molecular and hydrogen line emissinos (Kaifu
et al. 1972; Scoville 1972; Tsuboi 1988; Sofue 1990, 1991).
The molecular expanding ring of 200 pc radius  is the most tyipical example,
which has a radius of 200 pc and is expanding at a velocity of 50 \kms.
The tangential points to  the 200-pc ring on the $l-v$ diagram appear at
$l=1\Deg.6$ and at $l=-1\Deg.2$.
Cross sections of the ring in CO emission at these longitudes indicates that
they compose walls of 150 pc height perpendicular to the galactic plane with a
thickness of about 20 pc.
Namely, the expanding ring is a cylinder about 150 pc long (Sofue 1989).
The kinetic energy of the expansion motion is of the order of $10^{54}$ ergs,
which is comparable to that involved in the galactic center lobe. The
coincidence between the energies in the 200-pc cylinder and the GCL is
consistent if the GCL will focus on the ring in the future, and the 200-pc ring
is the consequence of the shock focusing of a past GCL some $10^6$ y ago.

A cylindrical molecular ring has been also found in the starburst galaxy M82
(Nakai et al 1987). The "200-pc ring" in M82 is thought to be the consequence
of an energy release, while far more intense than that in our Galaxy center, by
successive explosions of supernovae in the central region of the dense
molecular disk, which is followed by a vertical wind of high-temperature gas as
well as the formation of an expanding ring and a cylinder. Existence of
vertical magnetic field is not known, while the intense nonthermal radio
emission suggests field strength as high as some 10 \microg.

A more number of cylindrical features in the molecular gas has been identified
by Tsuboi (1988), who called them ``barrels''.
Barrels are rings and loops on position-velocity diagrams in the galactic
center region with a typical diameter of 30 pc and expanding motion of a few 10
\kms.
A detailed comparative study of radio continuum and CO-line emissions in the
central few degrees has revealed many loops and shells of molecular gas which
surround HII regions in the nulclear disk and are expanding at a velocity of a
few 10 \kms (Sofue 1990).
Typical molecular loops (or shells) are found enclosing Sgr B, Sgr C and  Sgr
D. Molecular gas and continuum thermal sources (HII regions) appear to avoid
each other.

\sect{\it 4.4. Infalling-Clouds and ``Galactic Sprays''}

The energy required to produce many of the expanding shells and cylinders is
typically $10^{52} - 10^{54}$ ergs.
The energy source could be supernova explosions or explosion at the nucleus.
We here suggest an alternative possibility of external energy source to create
such vertical structures:
These structures could be attributed to infalling gaseous debris from the
companion galaxies, LMC and SMC.
According to the ram-pressure stripping-and-accretion model of gas clouds in a
companion galaxy by its parent's gaseous halo and disk (Sofue 1994),
interstellar clouds are stripped and accreted toward the disk of the larger
galaxy .
If we apply this idea to the Magellanic Clouds and the Galaxy, gas clouds in
LMC and SMC are stripped, which soon form the Magellanic Stream, and finally
infall toward the galactic disk.
If the clouds' orbit is retrograde with respect to the galactic rotation, the
clouds hit the nuclear disk, where their sprays would exhibit various vertical
ridges and peculiar kinematics.
The kinetic energy given to the nuclear disk by a collision of an infalling
cloud of mass of $10^6\Msun$, a typical GMC, is  $\sim 4\times 10^{53}$ erg for
an infalling velocity of 200 \kms.
This energy is enough to produce any of the vertical and dynamical (expanding)
phenomena as discussed in this article.

\v\v
\noi{\bf References}
\parskip=0pt
\vskip 5mm

\ref{Altenhoff,  W. J.,  Downes,  D.,  Pauls., T.,  Schraml,  J. 1979,  AA
Suppl,   35,  23}

\ref{Bally,  J.,  Stark,  A.A.,  Wilson, R.W.,  Henkel, C. 1987,  Ap.J. Suppl,
 65,  13}

\ref{Berkhuijsen, E.M.,  Beck, R.,  Gr{\"a}ve, R. 1987,  in {\it Interstellar
Magnetic Fields},  ed. R.Beck and R.Gr{\"a}ve (Springer Verlag,  Berlin),
p.38}

\ref{Brown, R.L,  Liszt, H.S. 1984,   ARAA,   22,  223}

\ref{Downes, D.,  Goss, W.M.,  Schwarz, U.J.,  Wouterloot, J.G.A. 1978,  AA
Suppl,  {\bf 35},  1}

\ref{Dulk, G.A.,  Slee, O.B. 1974,  Nature,   248,  33}

\ref{Duric, N.,  Seaquist, E.R.,  Crane, P.C.,  Bignell, R.C.,  Davis, L.E.,
1983,  Ap.J.L,   273,  L11}

\r Falcke, H., Mannheim, K., Biermann, P. L. 1993, AA, in press.

\ref{F{\"u}rst, E.,  Sofue, Y,  Reich, W. 1987,  AA, 191,  303}

\ref{Genzel, R.,  Townes, C, H 1987,  ARAA,  25,  377}

\ref{G{\"u}sten, R. 1989,  in {\it The Center of the  Galaxy},  ed. M. Morris
(\kluwer),  p.89}

\ref{Haslam, C.G.T.,  Salter, C.J.,  Stoffel, H.,  Wilson, W.E.,  1982,  AA
Suppl, 47,  1}

\r Haynes, R. F. ,, Stewart, R. T., Gray, A. D., Reich, W., Reich, P., Mebold,
U. 1992, AA

\ref{Inoue, M.,  Takahashi, T.,  Tabara, H.,  Kato, T.,  Tsuboi, M.  1984,
PASJ, 36,  633}

\ref{Kaifu, N.,  Kato, T.,  Iguchi, T. 1972,  Nature,  238,  105}

\ref{Kapitzky, J.E.,  Dent, W.A. 1974,  Ap.J., 188,  27}

\r Krichbaum, T.  P., Zensus, J. A., Witzel, A>, Mezger, P. G., Standlke, K.,
et al. 1993, AA, 274, L37.

\ref{LaRosa, T.N.,  Kassim, N.E. 1985,  Ap.J.L,  299,  L13}

\ref{Liszt, H.S. 1985,  Ap.J.L, 293,  L65}

\ref{Little, A.G. 1974,  in {\it Galactic Radio Astronomy,  IAU Symp. No.60},
ed. F.J.Kerr and S.C.Simonson III (D.Reidel,  Dordrecht),  p.349}

\ref{Lo, K.Y. 1986a,   Science, 233,  1394}

\ref{Lo, K.Y. 1986b,  PASP,  98,  179}

\ref{Mezger, P.G.,  Pauls, T. 1979,  in {\it The Large-scale characteristics of
the lGalaxy,  IAU Symp. No.84},  ed. W.B.Burton (D.Reidel,  Drodrecht),  p.357}

\ref{Mills, b.Y.,  Drinkwater, M.J. 1984,  Austr.J.Phys.,   5,  43}

\ref{Morris, M. 1993,  in this issue}

\ref{Morris, M.,  Yusef-Zadeh, F. 1985,  AJ,  90,  2511}

\ref{Pauls, T.,  Downes, D.,  Mezger, P.G.,  Churchwell, W. 1976,  AA, 46,
407}

\r Predehl, P. 1993, in this issue.

\ref{Reich, W. 1993,  in this issue}

\ref{Reich, W.,  F{\"u}rst, E.,  Steffen, P.,  Reif, K.,  Haslam, C.G.T. 1984,
 AA Suppl, 58,  197}

\r Reich, W., Reich, P., F{\"u}urst, E. 1990, AA Suppl.

\ref{Reich, W.,  Sofue, Y.,  F{\"u}rst, E. 1987,  PASJ, 39,  573}

\ref{Reich, W.,  Sofue, Y.,  Wielebinski, R.,  Seiradakis, J.H. 1988, 191, 303}

\ref{Scoville, N.Z. 1972, ApJ.L., 175,  L127}

\ref{Seiradakis, J.H.,  Lasenby, A.N.,  Yusef-Zadeh, F.,  Wielebinski, R.,
Klein, U. 1985,  Nature, 17,  697}

\ref{Seiradakis, J.H., Reich, W.,  Wielebinski, R., Lasenby, A. N.,
Yusef-Zadeh, F. 1989, AA Suppl. 81, 291.

\ref{Serabyn, E.,  G{\"u}sten, R. 1986,  AA, 161,  334}

\ref{Serabyn, E.,  Lacy, J.H. 1985,  Ap.J, 293,  445}

\ref{Shibata, K.,  Uchida, Y. 1987,  PASJ, 39,  559}

\ref{Sofue, Y. 1984,  PASJ,  36,  539}

\ref{Sofue, Y. 1985,  PASJ,  37,  697}

\r \so\ 1989,  Ap. Let.  Commun.,  28, 1.

\r \so\ 1990, PASJ, 42, 827.

\ref{Sofue, Y. 1989,  in in {\it The Center of the Galaxy} (ed. M. Morris,
\kluwer), p.213.

\r Sofue, Y. 1994, ApJ, March, in press.

\ref{Sofue, Y.,  Fujimoto, M. 1987,  Ap.J.L, 319,  L73}

\ref{Sofue, Y.,  Fujimoto, M. 1987, PASJ, 39, 843.

\ref{Sofue, Y.,  Handa, T. 1984,  Nature,  310,  568}

\ref{Sofue, Y.,  Inoue, M.,  Handa, T.,  Tsuboi, M.,  Hirabayashi, H.,
Morimoto, M.,  Akabane, K. 1986,  PASJ, 38,  483}

\r Sofue, Y., Murata, Y., Reich, W. 1992, PASJ, 44, 367.

\ref{Sofue, Y.,  Reich, W.,  Inoue, M.,  Seiradakis, J.H. 1987,  PASJ, 39, 359}

\ref{Sofue, Y.,  Reich, W.,  Reich.P.,  1988, Ap.J.L, 341, L47.

\ref{Tsuboi, M. 1993,  in this issue}

\ref{Tsuboi, M.,  Inoue, M.,  Handa, T.,  Tabara, H.,  Kato, T.,  Sofue, Y.,
Kaifu, N. 1986,  AJ, 92,  818}

\r Tsuboi, M., Handa, T., Inoue, M., Ukita, N., Takano, T. 1988, PASJ, 40, 665.

\r Uchida, K. I., Morris, M. R., Serabyn, E., and Bally, J.  1993, ApJ. in
press.

\ref{Uchida, Y.,  Shibata, K. 1986, PASJ, 38,  }

\ref{Uchida, Y.,  Shibata, K.,  Sofue, Y. 1985, Nature, 317, 699}

\r van Albada, R. D. 1980, AA Suppl. 39, 283.

\ref{Yusef-Zadeh, F. 1989, in {\it The Center of the Galaxy} (ed. M. Morris,
\kluwer), p.243.

\ref{Yusef-Zadeh, F.,  Morris, M. 1986,  Ph.D. Theis, Columbia University}

\ref{Yusef-Zadeh, F.,  Morris, M. 1988,  ApJ, 326,  574}

\ref{Yusef-Zadeh, F.,  Morris, M.,  Chance, D. 1984, Nature,  {\it Nature},
310, 557}

\ref{Yusef-Zadeh, F.,  Morris, M.,  Slee, O.B.,  Nelson, G.J. 1986,  ApJ, 310,
689}

\ref{Yusef-Zadeh, F.,  Morris, M.,  Slee, O.B.,  Nelson, G.J. 1986,  ApJL, 300,
L47}

\endpage
\parskip 10pt
\centerline{\bf Figure Captions}
\v\v

\hsize=50truemm

\noindent Fig. 1: (a) 10 GHz continuum map of the Galactic center $3\Deg \times
2.5\Deg$ region. (b) 5 GHz VLA filaments of the radio Arc. Tick interval is
1$'$. (Yusef-Zadeh 1986)

\hsize=126truemm

\v
\noindent Fig. 2: (a) North Polar Spur at 408 MHz. $180\Deg \times 180\Deg$
area is shown. (background subtracted; from Haslam et al 1982).

(b) NGC 3079 at 1420 MHz (Duric et al 1983).

(c) NGC 4258 at 1420 MHz (van Albada 1982).
Galactic planes are horizontal.

\bye